\documentclass[conference]{IEEEtran}
\IEEEoverridecommandlockouts

\usepackage{cite}
\usepackage{url} 
\usepackage{amsmath,amssymb,amsfonts}
\usepackage{algorithmic}
\usepackage{graphicx}
\usepackage{subcaption} 
\usepackage{textcomp}
\usepackage{xcolor}
\usepackage{quantikz}
\usepackage{tikz}
\usepackage{amsmath}
\usepackage{mathrsfs, comment}
\usepackage{array}
\usepackage{adjustbox}
\usepackage{color,soul}
\usepackage[tt=false, type1=true]{libertine}
\usepackage[utf8]{inputenc}
\usepackage{algorithm,multirow}

\usepackage{cite}
\usepackage{amsmath,amssymb,amsfonts}
\usepackage{algorithmic}
\usepackage{graphicx}
\usepackage{textcomp}
\usepackage{xcolor}
\usepackage{algorithm}
\usepackage{algorithmic}
\usepackage{quantikz}
\usepackage{balance} 

\usepackage{comment}
\usepackage{multirow}

\usepackage{url} 
\usepackage{hyperref}

\usepackage[tableposition=bottom]{caption}

\newcommand{\sol}{DisMap}
\newcommand{\link}{Entanglement link}

\def\BibTeX{{\rm B\kern-.05em{\sc i\kern-.025em b}\kern-.08em
    T\kern-.1667em\lower.7ex\hbox{E}\kern-.125emX}}
    
\begin{document}

\title{Hardware-aware Circuit Cutting and Distributed Qubit Mapping for Connected Quantum Systems}

\author{Zefan Du, Yanni Li, Zijian Mo, Wenqi Wei, Juntao Chen, Rajkumar Buyya and Ying Mao
\IEEEcompsocitemizethanks{
\IEEEcompsocthanksitem Zefan Du, Yanni Li, Wenqi Wei, Juntao Chen and Ying Mao are with the Department of Computer and Information Science, Fordham University, New York City. E-mail: \{zdu19, yl32, wwei23, jchen504, ymao41\}@fordham.edu
\IEEEcompsocthanksitem Zijian Mo is with BAE Systems, USA 
\IEEEcompsocthanksitem Rajkumar Buyya is with University of Melbourne, Australia. E-mail: rbuyya@unimelb.edu.au  
}}


\maketitle

\begin{abstract}
Quantum computing offers unparalleled computational capabilities but faces significant challenges, including limited qubit counts, diverse hardware topologies, and dynamic noise/error rates, which hinder scalability and reliability. Distributed quantum computing, particularly chip-to-chip connections, has emerged as a solution by interconnecting multiple processors to collaboratively execute large circuits. While hardware advancements, such as IBM's Quantum Flamingo~\cite{ibmQuantumDelivers}, focus on improving inter-chip fidelity, limited research addresses efficient circuit cutting and qubit mapping in distributed systems. This project introduces \sol, a self-adaptive, hardware-aware framework for chip-to-chip distributed quantum systems. \sol~analyzes qubit noise and error rates to construct a virtual system topology, guiding circuit partitioning and distributed qubit mapping to minimize SWAP overhead and enhance fidelity. Implemented with IBM Qiskit and compared with the state-of-the-arts, \sol~achieves up to a 20.8\% improvement in fidelity and reduces SWAP overhead by as much as 80.2\%, demonstrating scalability and effectiveness in extensive evaluations on real quantum hardware topologies.
\end{abstract}

\section{Introduction}

Quantum computing promises unparalleled computational capabilities by harnessing the principles of quantum mechanics and pushing the boundaries in many fields~\cite{stein2021hybrid, mu2022iterative, sheng2023quantum, baheri2022pinpointing, ruan2022vacsen, d2023distributed, ruan2023venus, jiang2024resource, ruan2023quantumeyes, 9605352, stein2022quclassi, stein2022qucnn, fomichev2024initial}. Companies like IBM-Q, Google Quantum AI, IonQ, and Quantinuum have made significant advances in quantum hardware, achieving milestones in qubit count and gate fidelity. However, fundamental challenges persist: current quantum machines have limited qubit counts, far below practical requirements, and are constrained by diverse topologies, connectivity limitations, and dynamic noise/error rates. These obstacles present critical challenges to scalability and reliability, motivating intense research on both hardware and software fronts.

On the hardware side, distributed quantum computing has emerged as a promising solution to overcome the limitations of monolithic devices. This approach interconnects multiple quantum processors, enabling them to collaboratively solve complex problems. Distributed systems can be categorized into two approaches: quantum networks and chip-to-chip connections. Quantum networks link processors over long distances using entanglement and quantum communication protocols, enabling applications like quantum cryptography and distributed algorithms. Chip-to-chip connections, on the other hand, integrate multiple quantum chips within a single system or across short distances, allowing direct qubit interactions. Particularly in superconducting architectures, where single-chip qubit counts are limited by physical constraints, reliable inter-chip connections significantly enhance computational capacity and enable larger, more complex circuits.
On the software side, quantum compilers address the challenges of diverse hardware topologies and dynamic qubit characteristics by mapping logical qubits to physical qubits. These techniques aim to optimize execution by minimizing noise and enhancing gate fidelity. For instance, IBM has developed techniques to optimize qubit mapping~\cite{ibmTranspilerlatest, li2019noise} and routing in their superconducting systems, while IonQ emphasizes remote Ion-Ion entanglement to scale trapped-ion architectures \cite{wright2019benchmarking, ionqIonQDemonstrates}. Additionally, various benchmarks have been proposed to studied their performance from various perspectives~\cite{kan2024benchmarking, li2023qasmbench, chen2023benchmarking}.

Despite these efforts, chip-to-chip distributed quantum computing has primarily focused on hardware improvements for enhancing the fidelity of remote operations, such as IBM Quantum Flamingo~\cite{ibmQuantumDelivers} that connects two Heron R2 chips with four connectors measuring up to a meter long. Limited research addresses the critical task of mapping large circuits onto distributed systems to utilize qubit resources effectively. Conversely, existing qubit mapping frameworks and compilers primarily optimize for individual machines with fixed topologies, leaving distributed environments largely unexplored.



This project proposes \sol~that bridges this gap by addressing the distributed qubit mapping problem for chip-to-chip connected quantum systems under superconducting architectures. Given multiple quantum processors, each with a fixed topology, we analyze qubit noise levels and error rates to identify optimal candidate qubits for establishing connections. These connections are used to virtually integrate the individual topologies into a unified system. This unified virtual system topology guides the partitioning of large circuits into smaller subcircuits, which are then mapped to individual processors. Our approach minimizes execution costs and improves fidelity, offering an effective framework for distributed quantum computing. \sol~considers multiple hardware topologies and their noise levels to the guide circuit cutting and distributed qubit mapping. Our contributions are as follows:

\begin{itemize}
    \item We introduce circuit cutting and qubit mapping in chip-to-chip connected distributed quantum systems, aiming to enable large circuit execution by leveraging multiple machines and reducing SWAP overhead.
    
    \item We propose \sol, a self-adaptive, hardware-aware framework. It identifies best qubit pairs to construct a virtually connected system based on qubit's noise levels. The virtual system topology is utlized to guide circuit cutting aiming to increase the fidelity. 
    
    \item We implement \sol~using IBM Qiskit and evaluate it on real quantum hardware topologies. Extensive experiments demonstrate that \sol~outperforms the state-of-the-art qubit mapping and circuit cutting solutions. \sol~ achieves up to a $20.8\%$ improvement in fidelity and reduces SWAP overhead by as much as $80.2\%$.
\end{itemize}

\section{Background and Motivation}
This section provides an overview of quantum circuit cutting, chip-to-chip distributed quantum computing, and noise challenges, followed by the motivation for this work.

\subsection{Quantum Circuit Cutting}

Quantum circuit cutting enables the decomposition of large circuits into smaller subcircuits that can be independently executed on current quantum hardware. This technique mitigates the limitations of devices with restricted qubit counts by partitioning circuits into manageable segments and reconstructing the results through classical post-processing~\cite{bravyi2016trading, peng2020simulating}. Two primary cutting methods are employed: wire cuts, which separate a qubit's trajectory between gates, and gate cuts, which distribute multi-qubit gates like CNOT gates across subcircuits~\cite{peng2020simulating}. These approaches reduce circuit complexity, making larger algorithms feasible. Wire cuts are particularly effective in hybrid quantum-classical workflows, where quantum hardware executes quantum operations and classical systems handle reconstruction. However, as circuit size and the number of cuts increase, the reconstruction overhead grows exponentially, posing scalability challenges.


\subsection{Chip-to-Chip Distributed Quantum Computing}

Chip-to-chip distributed quantum computing addresses the limitations of monolithic devices by interconnecting multiple quantum chips, expanding computational capacity and enabling larger quantum circuits. Central to this approach is entanglement, which facilitates quantum communication between chips~\cite{einstein1935can}. Entanglement-based connections, including remote gates, quantum teleportation, and Einstein-Podolsky-Rosen (EPR) pairs, are essential for distributed systems. EPR pairs, in particular, are widely used due to their simplicity and compatibility with key protocols. They establish entangled states between qubits on separate chips, enabling non-local gate operations and secure quantum information transfer. For instance, the maximally entangled state $\frac{1}{\sqrt{2}} (\ket{00} + \ket{11})$ underpins tasks like teleportation and entanglement swapping~\cite{andres2019automated}.

While chip-to-chip connected systems still face challenges such as hardware limitations, noise, decoherence, and photon loss, which impact entanglement fidelity and computational efficiency, advances in superconducting qubits are promising and have achieved local link success rates exceeding 90\% for short-range connections~\cite{song2019generation}. For example, IBM Quantum Flamingo~\cite{ibmQuantumDelivers} connects two Heron R2 chips with four connectors measuring up to a meter long and 3.5\% errors. 
However, photonic platforms and trapped-ion systems leverage error correction and entanglement purification to enhance robustness~\cite{wang2020integrated, luo2020trapped}. 

\subsection{Noises and Qubit Mapping}

Noise remains a fundamental challenge in quantum computing, as quantum states are highly sensitive to hardware imperfections and environmental disturbances. Common sources include qubit decoherence, gate errors, cross-talk, and inefficiencies in entanglement generation. These issues are particularly detrimental in circuits with numerous two-qubit operations, demanding robust error mitigation techniques~\cite{bravyi2016trading}. 

Gate errors result from inaccuracies in implementing quantum gates, with single-qubit errors caused by calibration imperfections and two-qubit errors stemming from complex qubit interactions~\cite{ballance2016high, barends2014superconducting}. Cross-talk noise arises when operations on one qubit inadvertently affect neighboring qubits due to electromagnetic coupling, particularly in densely packed arrays~\cite{krantz2019quantum, mckay2017efficient}. In distributed systems, entanglement-based connections introduce additional noise due to decoherence and photon loss, with optical systems experiencing reduced success rates over long distances. Noise mitigation techniques, including entanglement purification and quantum repeaters, are essential for maintaining fidelity in distributed systems.

Qubit mapping, the process of assigning logical qubits to physical qubits, directly influences circuit execution fidelity by optimizing qubit assignments to minimize noise and hardware constraints~\cite{ibm_qiskit}. In distributed systems, mapping must also account for inter-chip entangled links, like EPR pairs, ensuring efficient resource utilization and enhanced circuit performance.


\subsection{Motivation}


Existing approaches in circuit cutting~\cite{kan2024scalable, qiskit-cutting, lowe2023fast, brandhofer2023optimal, smith2023clifford, bechtold2023investigating} and qubit mapping~\cite{hua2023qasmtrans, khadirsharbiyani2023trim, zhang2021time, cheng2024robust, wang2024atomique, jin2024tetris, ren2024leveraging} focus primarily on single-device systems with fixed topologies. For example, Qiskit Addon-Cutting~\cite{qiskit-cutting} requires user-specified parameters, like the number of cuts and qubit constraints, which are challenging to optimize manually. FitCut~\cite{kan2024scalable} automates this process to fit hardware constraints but maximizing qubit capacity leaves little room for qubit mapping and it fails to account for noise during partitioning. Additionally, all existing qubit mapping techniques, such as IBM’s default compiler Sabre~\cite{sabre2019}, optimize for single-device environments, ignoring the complexities of multi-node distributed quantum systems.
In contrast, we address chip-to-chip connected quantum systems by integrating circuit cutting with distributed qubit mapping. Our approach optimizes circuit execution across diverse topologies while accounting for hardware configurations and noises, offering a robust solution for scalable quantum computing.

\section{Solution Design}
In this section, we present our problem setting and system designs with detailed algorithms. 

\subsection{Problem Setting and Objectives}

\begin{figure}
    \centering
    \includegraphics[width=0.9\linewidth]{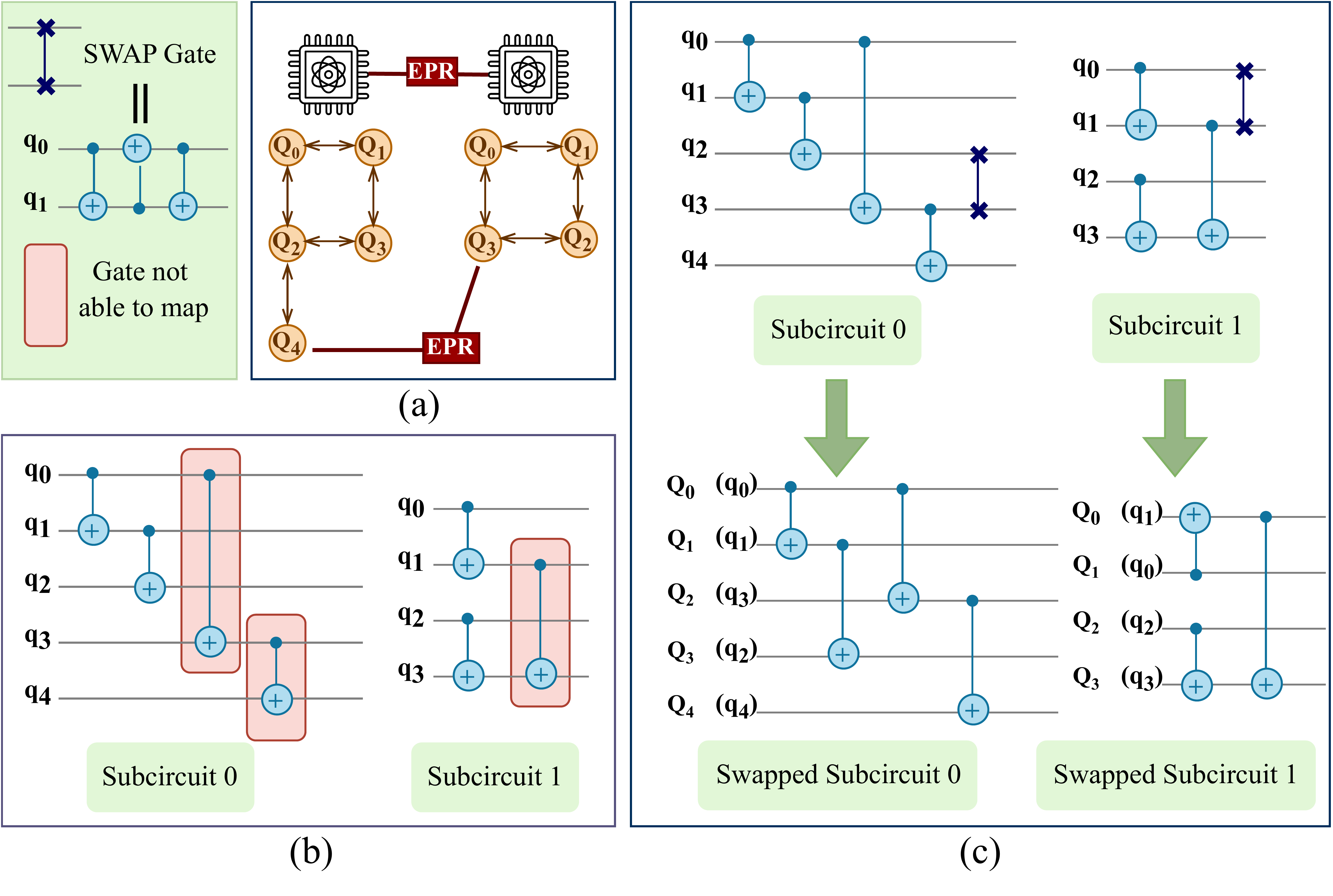}
    \caption{(a) Virtual System Topology ($VST$) with EPR between Q4 in worker-0, and Q3 in worker-1. (b) Subcircuits with gates that are not able to map into topology. (c) Qubit mapping with swapped subcircuits 0 \& 1 in worker 0 \& 1.}
    \label{fig:Example}
 \vspace{-0.15in}   
\end{figure}



We consider a distributed quantum system with \( n \) quantum workers \( W = \{w_1, \dots, w_n\} \), each characterized by a specific topology \( T = \{T_1, \dots, T_n\} \), qubit capacity \( Q = \{Q_1, \dots, Q_n\} \), and a noise model \( N = \{N_1, \dots, N_n\} \). Workers are dynamically interconnected via EPR pairs, denoted by \( E \), with an associated success rate \( SR \). While EPR pairs enhance connectivity and eliminate reconstruction in circuit cutting, they also introduce additional noise, which can degrade overall fidelity. In such a chip-to-chip connected system, a Virtual System Topology, \( VST \), is a graph such that multiple workers' topologies are connected via EPR pairs. A \( VST \) is illustrated in Figure~\ref{fig:Example}(a). After partitioning the circuit into subcircuits, directly placing a subcircuit onto a node by mapping subcircuit's logical qubits to hardware's physical qubits is not feasible due to unreachable two-qubit gate connections, as shown in Figure~\ref{fig:Example}(b). To address this, SWAP operations are employed to adjust the positions of logical qubits within the topology. This process produces a swapped subcircuit that fits within the Virtual System Topology in Figure~\ref{fig:Example}(c).


Given an input circuit \( C \) and a set of quantum workers, the problem involves partitioning \( C \) into \( m \) subcircuits \( S = \{s_1, \dots, s_m\} \) and mapping them to workers, considering the Virtual System Topology, \( VST \). Our approach addresses three key objectives:
(1) Identifying Entanglement Qubits: Select physical qubits for EPR pair connections \((q_k, q_l)\) between workers \( w_i \) and \( w_j \). The chosen pairs update the system topology \( VST \), incorporating their connectivity and noise profiles.
(2) Partitioning Circuits: Divide the input circuit \( C \) into subcircuits \( S \) based on \( VST \). This process prioritizes low-noise qubits within each worker \( w_i \), leveraging noise models \( N \) to optimize circuit execution.
(3) Mapping Subcircuits: Assign subcircuits \( \{s_1, \dots, s_m\} \) to workers, ensuring alignment with the physical topology and EPR connections. This step minimizes SWAP operations and operational noise by considering both intra- and inter-worker qubit allocation.

With chip-to-chip connected distributed quantum systems, we focus on two primary objectives. (1) {Maximizing Fidelity:} Select low-noise qubits and minimize noise introduced by quantum connections. This ensures high fidelity in distributed quantum computations.  (2) {Minimizing SWAP Overhead:} Optimize qubit allocation within and across workers to reduce SWAP operations, which are both resource-intensive and noise-prone. The total SWAP overhead is given by:

\begin{equation}
    SO =\sum_{i=1}^{m} SO_i = \sum_{i=1}^{m} \sum_{j=1}^n {\text{SWAPs on } W_j}
    \label{equ:swap}
\end{equation}
where \(SO_i\) represents the SWAP count for subcircuit \(s_i\), and \(W_j\) denotes the worker to which subcircuit \(s_i\) is assigned.



These objectives guide the selection of entangled qubit pairs, circuit partitioning, and subcircuit mapping. By balancing these factors, \sol~achieves efficient, high-fidelity execution of large quantum circuits in distributed systems.

\subsection{System Overview}

\begin{figure*}[htbp]
    \centering
    \includegraphics[width=0.95\linewidth]{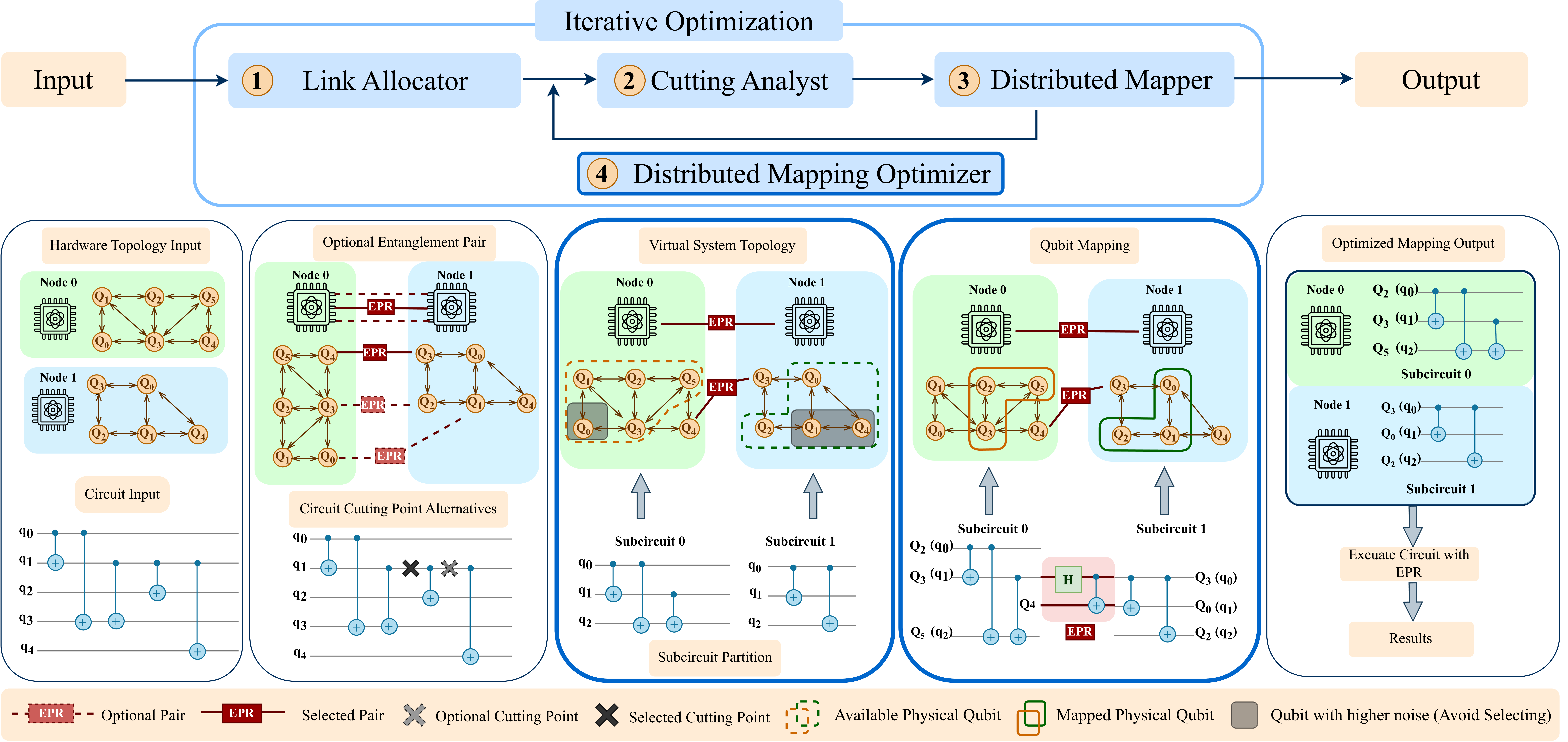}
    \caption{\sol~System Overview}
    \label{fig:overview}
\end{figure*}

\sol~ is a self-adaptive chip-to-chip distributed quantum computing platform designed to optimize the execution of large quantum circuits. It accepts two key inputs: (1) a client-provided logical circuit in formats such as Qiskit, Pytket, or PennyLane code, and (2) administrator-provided hardware specifications, including the number of worker nodes, their topologies, qubit capacities, and noise characteristics. Given these inputs, \sol~ operates under two scenarios: the circuit can be fully mapped onto a single quantum device within the system, or the circuit is too large to fit onto any single device. Existing qubit mapping techniques and quantum compilers primarily address Scenario 1. However, as input circuit size increases, these approaches offer limited optimization flexibility due to hardware constraints. In Scenario 2, which is more common in chip-to-chip connected distributed quantum systems, circuit cutting becomes essential. By partitioning the circuit and distributing subcircuits across multiple quantum workers, \sol~enables the execution of large quantum circuits without reconstruction. Key challenges in this scenario include identifying the optimal qubits within each worker for forming EPR links, partitioning the circuit effectively, and balancing the trade-off between maintaining high fidelity and minimizing SWAP overhead.

Figure~\ref{fig:overview} illustrates the \sol~ system overview. With hardware specifications and a logical circuit as inputs, \sol~ begins with {\em Step 1: Link Allocator}, selecting physical entanglement pairs (e.g., EPR connections) between two workers. Each pair involves one qubit from each worker, forming the Virtual System Topology ($VST$). In {\em Step 2: Cutting Analyst}, \sol~ analyzes qubit noise levels to make every effort to avoid selecting high-noise qubits (e.g., gray-shaded qubits), however, it is still possible to select that such qubits for partition. These subcircuits are pariationed and assigned to specific workers based on the $VST$. {\em Step 3: Distributed Mapper} maps the logical qubits of subcircuits to physical qubits according to the updated $VST$. For instance, subcircuits 0 and 1 may be assigned to Node 0 and Node 1, respectively. {\em Step 4: Distributed Mapping Optimizer} minimizes SWAP overhead by selecting optimal mappings between logical and physical qubits for each subcircuit, ensuring efficient hardware-aware partitioning. Step 1 to 4 is an iterative optimization process. \sol~ iterates through potential entanglement pair configurations to identify the best setup that balances circuit performance and resource constraints. The process concludes with the Output, where subcircuits are distributed to connected quantum workers, and the final results are generated.

\begin{algorithm}
\caption{\sol~Function Definitions}
\label{alg:FC}
\begin{algorithmic}[1]
\STATE \textbf{Function: ConVST}($T, EP_s$)
    \STATE\quad \textbf{FOR} {$(q_k\_w_i, q_l\_w_j)$ in $EP_s$}
            \STATE\quad\quad $VST \gets T\_w_i + T\_w_j + (q_k\_w_i, q_l\_w_j)$
    \STATE \quad \textbf{Return} $VST$, Virtual System Topology
    
\item[] 
\STATE \textbf{Function: TopoCutter}($C$, $VST$)
    \STATE\quad $SQ\_min \gets C.qubit()$
    \STATE\quad $SQ\_max \gets VST.qubit()$
    \STATE\quad\quad \textbf{FOR} {$sq$ in $[SQ\_min,SQ\_max]$}
        \STATE \quad\quad\quad $VST\_curr \gets (sq, VST)$
        \STATE \quad\quad\quad $S \gets (C, VST\_curr)$
    \STATE \quad \textbf{Return} $S$ subcircuits

\item[] 
\STATE \textbf{Function SubMapper}($S$, $VST$)
    \STATE \quad $MS = [~]$
    \STATE \quad \textbf{FOR} {$s_i$ in $S$}
        \STATE \quad\quad $ms_i:(s_i:W_j) \gets$ Transpiler($s_i$ in $VST$)
        \STATE \quad\quad $SO \gets sum(s_i:W_j)$ from Equation~\ref{equ:swap}  
        \STATE \quad\quad $MS.append(ms_i)$
    \STATE\quad \textbf{Return} $MS$ Mapped subcircuit, $SO$ SWAP Overhead
\end{algorithmic}
\end{algorithm}

\begin{algorithm}
\caption{\sol~Iterative Optimization}
\label{alg:FCN}
\begin{algorithmic}[1]
\STATE \textbf{Input:} Circuit: $C$, System configurations: Topology $T$ on workers $W$ with Noise $N$
\STATE $EP_s: [[(q_k\_w_0, q_l\_w_1):SR_0, \dots,(q_k\_w_i, q_l\_w_j):SR_n],\dots] \gets (T, W, N)$
\STATE $SO \gets Inf$
\STATE \textbf{FOR} {$EP_s\_i$ \textbf{in} $EP_s$}
    \STATE\quad $VST \gets$ ConVST($T, EP_s\_i$)
    \STATE\quad $S \gets$ TopoCutter($C, VST$)
    \STATE\quad $MS\_curr, SO\_curr \gets$ SubMapper($S, VST$)
    \STATE \quad \textbf{IF} $SO\_curr < SO$
        \STATE \quad\quad $SO \gets SO\_curr$
        \STATE \quad\quad $MS \gets MS\_curr$
\STATE  $R, F \gets$ Execute $MS$ on $VST$
\STATE \textbf{Return} Result $R$, SWAP Overhead $SO$, Fidelity $F$
\end{algorithmic}
\end{algorithm}

\subsection{Algorithms Design}



\sol~optimizes circuit cutting and subcircuit qubit mapping by considering hardware topologies and noise levels. Algorithm~\ref{alg:FC} defines three core functions: Constructing Virtual System Topology (ConVST), Topology-based Cutter (TopoCutter), and Subcircuit Qubit Mapper (SubMapper).
The ConVST function generates the Virtual System Topology (VST) by analyzing the topologies \(T\) of all workers and selecting the lowest-noise \link~pairs between them. The resulting VST captures the qubit layout and inter-worker connectivity. The TopoCutter function determines the range of available qubits, from \(SQ\_min\) (the number of qubits in the circuit \(C\)) to \(SQ\_max\) (the total qubits in the VST), and partitions the circuit \(C\) into subcircuits \(S\) based on the updated VST. The SubMapper function maps the subcircuits \(S\) onto the VST using a customizable transpiler, producing the mapped subcircuits (\(MS\)) and calculating the total SWAP overhead (\(SO\)) by Equation~\ref{equ:swap}.

Building on these functions, Algorithm~\ref{alg:FCN} introduces an iterative optimization framework. Given an input circuit \(C\) and a system configuration that includes workers (\(W\)), their topologies (\(T\)), and noise profiles (\(N\)), the algorithm identifies the lowest-noise qubits in each worker and generates a list of physical entanglement pairs (\(EP_s\)) with success rates (\(SR\)). The SWAP overhead (\(SO\)) is initialized to infinity, ensuring that each valid configuration improves upon the prior result. The algorithm iterates over all potential \link pair candidates in \(EP_s\). In each iteration, it first constructs the VST by invoking ConVST with the selected \link pairs. Once the VST is generated, the TopoCutter function partitions the circuit \(C\) into subcircuits \(S\). The subcircuits are then mapped to the VST using SubMapper, which returns the mapped subcircuits (\(MS\)) and the current SWAP overhead (\(SO\_curr\)) by equation~\ref{equ:swap}. If \(SO\_curr\) is less than the previously recorded value (\(SO\)), the algorithm updates \(SO\) and \(MS\) accordingly.
After completing all iterations, the algorithm executes \(MS\) on the VST, resulting in the final circuit output (\(R\)), the total SWAP overhead (\(SO\)), and the fidelity (\(F\)). This iterative process ensures the optimal selection of \link connections, effectively minimizing SWAP overhead while maximizing fidelity in distributed quantum computations.

\section{Performance Evaluation}

We evaluate the performance of \sol~ with a focus on fidelity and SWAP operation overhead in distributed quantum systems with varied topologies. 

\subsection{Implementation and Evaluation Settings}

\sol~ is implemented in Python 3.10 using Qiskit 1.2 for quantum operations. Due to the lack of hardware access to chip-to-chip connections, we used IBM-Q noisy emulators~\cite{ibmFake_providerlatest} derived from real quantum devices, such as AlmadenV2, Auckland, TorontoV2, SydneyV2, and MontrealV2. Classical computations were performed on a Google Cloud e2-highmem-16 instance with AMD Rome x86/64 processors. The evaluation setup consisted of a simulated chip-to-chip distributed system with 3 or 4 workers, each configured with either 20 or 27 qubits. Dynamic EPR pairs (\link) were used for inter-node connections, with 1 to 2 \link~per system, enabling analysis across different connectivity scenarios. Additionally, the baseline transpiler used in SubMapper can employ any technique. For the following experiments, we selected IBM-Q's default transpiler, Sabre\cite{sabre2019}. Besides, FitCut~\cite{kan2024scalable} is used to generate our subcircuits under our VST configurations. This design makes \sol~compatible with existing circuit cutting and qubit mapping techniques. 

Our input circuits contain various complexity and qubit counts (18 to 60), including popular algorithms and randomly generated circuits. Workloads included the Bernstein-Vazirani (BV) algorithm for hidden bit string determination~\cite{bernstein1993quantum}, 
binary addition circuits (ADDER)~\cite{zhang2023characterizing}, 
Hardware Efficient Ansatz (HWEA) commonly used in variational algorithms~\cite{kandala2017hardware}, and the Quantum Approximate Optimization Algorithm (QAOA) for combinatorial problems~\cite{farhi2014quantum}. These workloads enable a comprehensive evaluation of \sol.

Based on these workloads, we evaluate two key metrics: fidelity, which measures the quality of execution, and the total number of SWAPs, which reflects the efficiency of qubit mappings. Additionally, we report the time overhead for our experiments. As baselines, we compare \sol~with IBM Qiskit's default compiler, {\bf Sabre}. Since Sabre does not support distributed environments, we implement {\bf Sabre-VST}, which manually provides it with a virtually connected topology. Furthermore, because Sabre-VST lacks circuit-cutting capabilities, we also implement {\bf Sabre-Addon-Cutting}, which combines Sabre-VST with Qiskit's Addon-Cutting~\cite{qiskit-cutting}. Since Qiskit's Addon-Cutting requires qubit constraints and a number of cuts as input, we manually set constraints to the minimum qubit capacity and the number of cuts to the worker count.

\begin{figure*}[htbp]
    \centering
    \includegraphics[width=0.92\linewidth]{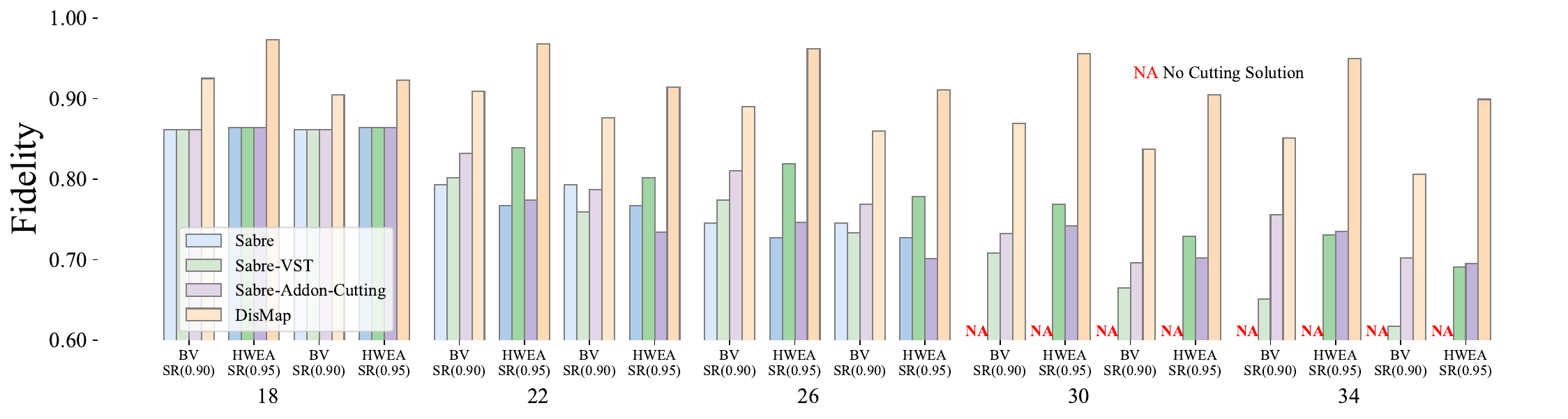}
    \caption{A 3-worker system with AlmadenV2, Auckland and TorontoV2 and input circuit sizes from 18 to 34. }
    \label{fig:F}
    ~
    \centering
    \includegraphics[width=0.92\linewidth]{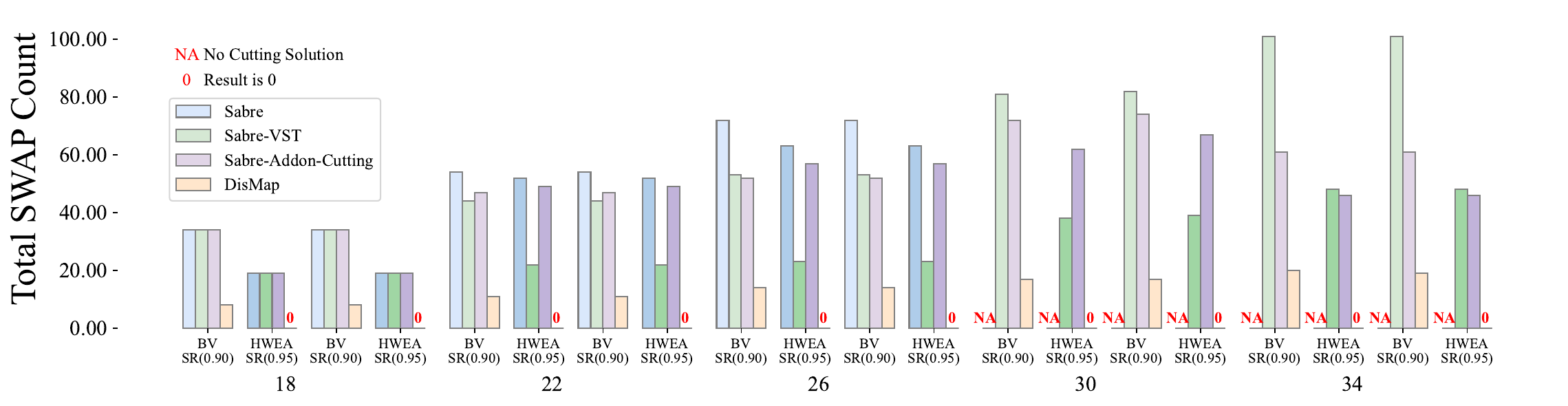}
    \caption{A 3-worker system with AlmadenV2, Auckland and TorontoV2 and input circuit sizes from 18 to 34.}
    \label{fig:S}
\end{figure*}


\subsection{Experiment Results}
We first construct a 3-worker distributed quantum system with  IBM-Q Emulators: AlmadenV2, Auckland and TorontoV2, that have a qubit capacity 20, 27 and 27, respectively. \sol~ is evaluated with two link construction Success Rate (SR) of 0.90 and 0.95 based on the hardware data in the literature. We have two types of circuits, BV and HWEA, with qubit counts ranging from 18 to 34.

Figure~\ref{fig:F} presents the fidelity. Obliviously, \sol~consistently demonstrates the highest fidelity across all experiments. Noticeably, when circuit size is smaller than an individual machine's capacity, circuit cutting is not required. However, our results demonstrate that \sol~can still benefit the fidelity. This is because, \sol~ can cut the circuit into smaller subcircuits and map them onto multiple machines. It provides more flexibility in qubit mapping phase that can avoid using low-quality qubits on a single machine. For example,  with a 18-qubit BV circuit and $SR = 0.95$, \sol~ records the highest fidelity of $97.3\%$, while others achieve $86.4\%$, representing an $10.9\%$ improvement. For BV and HWEA circuits with 18 qubits, Sabre, Sabre-VST, and Sabre-Addon-Cutting produce identical results since the circuit can be executed on a single worker and Sabre-Addon-Cutting only cut the circuit when the circuit cannot fit onto an individual machine. For circuit size 30 and 34, Sabre is not able to complete qubit mapping due to the hardware capacity limits and lack of cutting supports (i.e., represented by NA on the Figure~\ref{fig:F}). The largest improvement of fidelity achieved is a 34-qubit HWEA with $SR = 0.95$, where Sabre-VST records 69.1\% and \sol~ is 89.9\% , a 20.8\% increase.

Figure~\ref{fig:S} shows the total SWAP count for the same experiment. Compared to other methods, \sol~consistently achieves the lowest total SWAP count. By partitioning subcircuits onto lower-noise physical qubits, \sol~significantly reduces SWAP operations. For the 18-qubit BV circuit, \sol~records only 8 SWAPs, whereas the other three methods require $34$ SWAPs, representing a $76.5\%$ reduction. For all HWEA circuits, the SWAP number of \sol~reduce to $0$. \sol~remains consistently zero because the HWEA is specifically designed for the hardware. As a hardware-aware algorithm, \sol~partitions the circuit without requiring any additional SWAP operations. In contrast, the three Sabre-based algorithms map the circuit using qubits that approach the capacity of the workers, which necessitates additional SWAP operations. The maximum reduction is observed at the BV circuit with 34 qubits, where \sol~reduces the SWAP count from 101 (with Sabre-VST) to 20, representing an $80.2\%$ improvement. Similarly, Sabre fails to produce results because the circuit exceeds the capacity of a single node and they are represented by NA on the Figure~\ref{fig:S}. 


\begin{table}[h]
\centering
\caption{Total SWAPs and Time Cost}
\label{tab:time}
\resizebox{0.42\textwidth}{!}{%
\begin{tabular}{|c|c|c|c|c|c|}
\hline
\multicolumn{2}{|c|}{} & \multicolumn{2}{c|}{\textbf{Sabre-Addon-Cutting}} & \multicolumn{2}{c|}{\textbf{DisMap}} \\ \hline

\textbf{Circuit Type} & \textbf{Qubits} & \textbf{SWAPs} & \textbf{Time(s)} & \textbf{SWAPs} & \textbf{Time(s)} \\ \hline
\multirow{7}{*}{QAOA} 
& 30  & 81 & 16.7 & 59 & 0.8 \\ \cline{2-6}
& 36  & 106 & 24.6 & 87 & 0.8 \\ \cline{2-6}
& 40  & 121 & 37.8 & 108 & 1.4\\ \cline{2-6}
& 46  & 132 & 28.6 & 114 & 2.2 \\ \cline{2-6}
& 50  & N/A & 300* & 156 & 3.2\\ \cline{2-6}
& 56  & N/A & 300* & 197 & 3.9 \\ \cline{2-6}
& 60  & N/A & 300* & 234 & 2.6\\ \hline
\multirow{7}{*}{Adder}
& 30  & 112 & 300* & 56 & 0.8 \\ \cline{2-6}
& 36  & 144 & 300* & 68 & 1.3 \\ \cline{2-6}
& 40  & 166 & 300* & 76 & 2.4 \\ \cline{2-6}
& 46  & 174 & 300* & 88 & 3.2 \\ \cline{2-6}
& 50  & 199 & 300* & 96 & 4.3 \\ \cline{2-6}
& 56  & 224 & 300* & 108 & 11.6 \\ \cline{2-6}
& 60  & 246 & 300* & 116 & 12.3 \\ \hline
\multicolumn{6}{|c|}{Note:N/A: No cutting solution; * : 300s API time limit}\\ \hline
\end{tabular}%
}
\end{table}




Next, we evaluate \sol~on larger circuits, specifically QAOA with depth 1 and ADDER, ranging from 30 to 60 qubits. The system size is also increased to 4 workers, each using AlmadenV2 with 20 qubits, and $SR=0.95$.

Table~\ref{tab:time} compares the SWAP counts and execution times of \sol~and Sabre-Addon-Cutting. \sol~consistently outperforms Sabre-Addon-Cutting in both metrics. For the 30-qubit ADDER circuit, \sol~achieves an execution time of 0.8s, 375x faster than Sabre-Addon-Cutting's 300s. Moreover, it reduces SWAP counts by 50\%, from 112 to 56. In the table, an asterisk (*) after the execution time indicates that Sabre-Addon-Cutting prematurely terminated due to its API limit, returning non-optimal results. Qiskit-Addon-Cutting imposes a 300-second API limit; if no optimal cutting points are found within this time, the API either returns the best available results or no results at all if nothing is found. For 50, 56 and 60-qubit QAOA,  Qiskit-Addon-Cutting cannot find any cutting points within the limit. Therefore, Sabre-Addon-Cutting cannot proceed to the qubit mapping phase and marked as N/A in the table. The inefficiency of Sabre-Addon-Cutting arises from its lack of optimizations, as it simply combines Sabre with Qiskit-Addon-Cutting. The Qiskit-Addon-Cutting is not scalable for distributed environments or systems with multiple qubit constraints~\cite{kan2024scalable}.  For the 30-qubit QAOA, \sol~is 21x faster than Sabre-Addon-Cutting, with execution times of 0.8s and 16.7s, respectively. \sol~also reduces SWAP counts by $27.2\%$, from 81 to 59. For all ADDER circuits, Sabre-Addon-Cutting consistently hits its API limit. Therefore, its returned cutting points are non-optimal. Comparing with it, \sol~partitions and maps 60-qubit ADDER in 12.3s, achieving a SWAP count of 116, a $52.8\%$ reduction from 246.

\section{Conclusion}
In this paper, we presented \sol, a self-adaptive framework for circuit cutting and distributed qubit mapping in chip-to-chip connected quantum systems. By constructing a virtual system topology based on hardware configurations and noises, \sol~partitions circuits and maps subcircuits to distributed quantum processors. Extensive experiments on real hardware topologies demonstrated \sol's ability to significantly enhance fidelity and reduce SWAP overhead, achieving improvements of up to 20.8\% and reductions of up to 80.2\%, respectively. These results underscore \sol's effectiveness in leveraging distributed quantum systems. The future work involves investigating other hardware architectures, such as trapped-ion and neutral-atom quantum platforms. 

\bibliographystyle{ieeetr}
\bibliography{refs}

\end{document}